\documentstyle[epsfig]{elsart} 
\begin{document}        
\begin{frontmatter}    
\title{\bf Energy calibration of the NaI(Tl) calorimeter of
           the SND detector using $e^+e^- \rightarrow e^+e^-$ events }   
\author{ M.N.Achasov\thanksref{adrr},}    
\author{ D.A.Bukin, }
\author{ T.V.Dimova, }
\author{ V.P.Druzhinin, }
\author{ V.B.Golubev, }
\author{ V.N.Ivanchenko, }      
\author{ A.A.Korol }
\thanks[adrr]{ E-mail: achasov@inp.nsk.su, FAX: +7(383-2)35-21-63}          
\address{ Institute of Nuclear Physics, 
          Novosibirsk,     
          630090,        
          Russia }        
\date{}    
\begin{abstract}
 Calibration of the three layer NaI(Tl) spherical calorimeter of the
 SND detector using electron -- positron scattering  events is described.
 Energy resolution of $5 \% \mathrm{(FWHM/2.36 )}$ for 500 MeV photons was
 achieved.
\end{abstract}
\end{frontmatter}

\section{Introduction}

\begin{figure}
\epsfig{figure=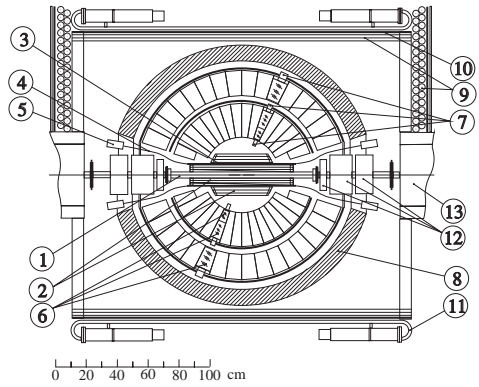,height=5cm}
\caption{SND detector, section along the beams; 1 --- beam pipe,
         2 --- drift chambers, 3 --- scintillation counters,
         4 --- light guides, 5 --- PMTs, 6 --- NaI(Tl) crystals,
         7 --- vacuum phototriodes, 8 --- iron absorber,
         9 --- streamer tubes, 10 --- 1cm iron plates,
         11 --- scintillation counters, 12 and 13 ---
         elements of collider magnetic system. }
\label{f1}
\end{figure}

 The SND is a general purpose nonmagnetic detector (~Fig.~\ref{f1}~)
 operating at VEPP-2M $e^+e^-$ collider in BINP (~Novosibirsk~) in the
 center of mass energy range of $0.2 \div 1.4$ GeV.
 \cite{SND,HADRON}. Experimental studies  include decays of
 $\rho, \omega, \phi$ mesons
 and nonresonance hadron production at low energies.
 Good energy resolution
 for photons in a wide energy range from $30$ to $700$ MeV is essential
 for suppression of
 background in reconstruction of intermediate $\pi^0$ and $\eta$ mesons and
 detection of photons emitted in radiative transitions between different
 quarkonium states.  Fast preliminary
 calibration of the calorimeter
 is based on cosmic muons \cite{CCC}. It provides reasonable
 energy resolution of  $5.5 \% \mathrm{(FWHM/2.36)}$ for 500 MeV photons,
 but to increase resolution to its highest value it is necessary to
 use experimental events with precisely known energies of final particles.
 In addition, such a process should have clear
 event topology, be well separated from background, and
 have large cross section. Potentially suitable processes are
 $e^+e^- \rightarrow e^+e^-, e^+e^- \rightarrow \gamma\gamma,
 e^+e^- \rightarrow \pi^+\pi^-, e^+e^- \rightarrow \mu^+\mu^-$,
 but given the VEPP-2M luminosity of
 $3 \cdot 10^{30} cm^{-2}s^{-1}$ at
 1 GeV, only $e^+e^- \rightarrow e^+e^-$ produces enough statistics for
 calibration of the calorimeter in a reasonably short time.

\section{The SND calorimeter}

\begin{figure}
\epsfig{figure=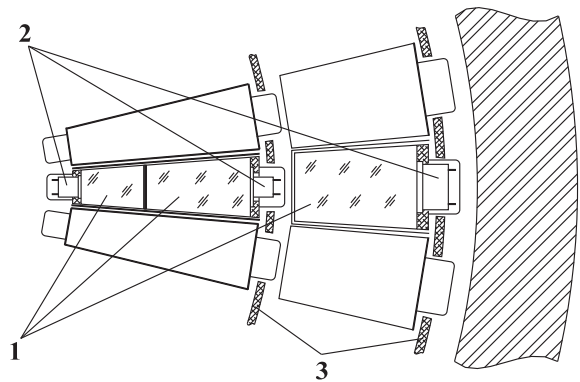,height=5cm}
\caption{NaI(Tl) crystals layout inside the calorimeter:
         1 --- NaI(Tl) crystals, 2 --- photodetectors (~vacuum phototriodes~),
         3 --- aluminum supporting hemispheres.}
\label{f2}
\end{figure}

 SND detector \cite{SND} (~Fig.~\ref{f1}) consists of a
 cylindrical drift chamber, calorimeter, and muon system.
 The three-layer NaI(Tl) spherical calorimeter, consisting of 1632
 individual counters
 (~Fig.~\ref{f2}) is a  main part
 of the detector. It was described in \cite{CCC}, so let us mention
 only some details necessary for description of the calibration
 procedure.

 Calorimeter solid  angle coverage in a spherical coordinate system 
 with $Z$ axis directed along the electron beam
 is $18^\circ \leq \theta \leq 162^\circ$
 and $0^\circ \leq \phi \leq 360^\circ$. Calorimeter is logically
 divided into two parts: ``small'' angles
 $18^\circ \leq \theta \leq 36^\circ$ and
 $144^\circ \leq \theta \leq 162^\circ$, and ``large'' angles:
 $36^\circ \leq \theta \leq 144^\circ$. The angular dimensions of crystals
 at ``large'' angles are $\Delta\phi = \Delta\theta=9^\circ$  and
 $\Delta\phi = 18^\circ, \Delta\theta=9^\circ$ at ``small'' angles.
 The calorimeter layers are enumerated starting from the interaction 
 point -- the first one is the nearest to the beam.

 The calorimeter energy resolutions for 500 MeV 
 electrons, photons, and muons after primary calibration using cosmic muons
 are shown in the Table~\ref{tab1}. While peak positions
 in Monte Carlo simulation and experiment
 agree at a $1 \%$ level, experimental resolutions
 are significantly worse than simulated ones.
 The possible explanations of the differences in simulated and experimental
 spectra could be attributed to
 instability of the detector electronics, systematic errors in
 the cosmic calibration procedure, and inadequate treatment of
 nonuniformity of light collection efficiency over the crystal volume in
 Monte Carlo simulation.

 Relative stability of calibration constants  in time \cite{CCC}, shows that 
 electronics and photodetectors instabilities do not contribute much
 into experimental resolutions. To eliminate systematic biases of
 cosmic calibration procedure, the OFF-LINE calibration based on
 $e^+e^- \rightarrow e^+e^-$ events was performed.

\section{The calibration method.}

 The calibration constants based on  $e^+e^-$ scattering events
 could be obtained by minimization of the expression:

\begin{equation}
 F(C_i) =  \sum\limits_j(\sum\limits_i U_{ij} \cdot C_i - E_0 )^2,
\end{equation}

 where $j$ -- event number, $i$ -- crystal number, $U_{ij}$ is an
 energy deposition in $i$th crystal and $j$th event, $E_0$ -- beam energy,
 $C_i$ -- calibration constant for $i$th crystal. Similar calibration
 procedure was implemented in CLEO II detector \cite{CLEO}, but layered
 structure of the SND calorimeter complicates the task due to large
 fluctuations of energy depositions in the calorimeter layers and energy
 dependence of longitudinal development of electromagnetic shower.
 Although r.m.s. of the energy deposition spectra in an individual crystals
 are about $100 \%$, the statistical accuracy of $C_i$ must be
 high, due to strong correlations between energy depositions in different
 crystals in an event. The drawbacks of such a direct method are that
 the calibration constants are dependent on the energy of the electrons
 and produce biased values of average energy depositions in calorimeter layers.

 To avoid such complications the
 SND calorimeter calibration is based on comparison of
 coefficients $C_{i}^{\mathrm{mc}}$ and
 $C_{i}^{\mathrm{exp}}$, obtained using simulated and experimental
 electrons of the same energy.

 After cosmic calibration the measured energy deposition in $i$th
 crystal $U_i$ can be written as
 $U_i = \epsilon_i \cdot E_i$, where $E_i$ is an actual energy deposition
 and difference of $\epsilon_i$ from unity, characterizes
 the systematic shift of cosmic calibration for $i$th crystal.
 To compensate for this shift it is necessary to find corresponding
 correction coefficient
 $C_{i}^{\mathrm{cal}} = 1 / \epsilon_i$.
 These calibration constants are obtained in the following way.
 First, the functions:

\begin{equation}
 F(C_i^{\mathrm{mc}}) = \sum\limits_{j}
 ( \sum\limits_{i} C_i^{\mathrm{mc}}E_{ij} - E_0 )^2~~~\mathrm{and}~~~
 F(C_i^{\mathrm{exp}}) = \sum\limits_{j}
 ( \sum\limits_{i} C_i^{\mathrm{exp}}U_{ij} - E_0 )^2,
\end{equation}

 are minimized over $C_i^{\mathrm{mc}}$ and
 $C_i^{\mathrm{exp}}$. Here $i$ is a crystal number, $j$ -- event number
 $E_{ij}$ -- energy depositions in crystals in Monte Carlo simulation,
 $U_{ij}$ -- measured energy depositions in experimental $e^+e^-$ events,
 $E_0$ -- beam energy  ( the same in experimental and simulated events ).
 The minimums of the functions are determined by the following conditions:

\begin{equation}
 \partial F(C_i) / \partial C_i = 0
\end{equation}

 Coefficients $C_i^{\mathrm{exp}}$ and $C_i^{\mathrm{mc}}$ are the 
 solutions of the set of simultaneous linear equations of the form
 $A\cdot c=b$, where $A$ is an  $n \times n$ matrix
  and its elements are   $A_{im} = \sum\limits_{j}E_{ij}E_{mj}$
 or $\sum\limits_{j}U_{ij}U_{mj}$, $c$ and $b$ are vectors of a dimension
 $n$ with elements  $c_{i} = C_i^{\mathrm{mc}}$ or $C_i^{exp}$
 and $b_i = E_0 \sum\limits_{j}E_{ij}$ or
 $E_0 \sum\limits_{j}U_{ij}$. Here $n=1680$ is a total
 number of crystals in calorimeter, indices $i$ and $m$ - crystals numbers,
 $j$ - the event number. The solutions of the two linear systems satisfy the
 following condition
 $C_{i}^{\mathrm{exp}} = C_{i}^{\mathrm{mc}} / \epsilon_{i}$, hence
 $C_{i}^{\mathrm{cal}} = C_{i}^{\mathrm{exp}} / C_{i}^{\mathrm{mc}}$. 

 The calibration constants $C_{i}^{\mathrm{cal}}$ obtained
 this way have high statistical accuracy, are independent of the
 energy of electrons, and do not produce biases in energy depositions
 in calorimeter layers.

 For calorimeter calibration $e^+e^- \rightarrow e^+e^-$ events were
 simulated. In order to save simulation time, angular distribution
 was set to uniform over the solid angle. The passage of electrons
 through the detector was simulated by means of the UNIMOD2 code
 \cite{UNIMOD}.

 Both experimental and simulated $e^+e^- \rightarrow e^+e^-$ events are
 selected according to the same criteria: only two particles must
 be detected, the total energy deposition in the calorimeter is greater than
 $1.2 \cdot E_0$ and acollinearity angle is less than 10 degrees.
 All crystals with energy depositions less than 5 MeV are discarded. Remaining
 crystals are put into calculation of the elements of matrix $A$ and vector
 $b$. When events processing is finished, the linear system is
 solved using SLAP2 \cite{SLAP} package, i.e.
 $C_{i}^{\mathrm{exp}}$ and $C_{i}^{\mathrm{mc}}$
 are obtained for all crystals and then $C_{i}^{\mathrm{cal}}$ are calculated.

\section{Events processing results.}

 This calibration procedure was used in the OFF-LINE processing of the
 data collected in $1996\div1997$ in the center of mass energy range
 $0.99 \div 1.04$ GeV \cite{SND}.

 To obtain $C_{i}^{\mathrm{mc}}$, 50000 simulated events with
 500 MeV electrons were processed, corresponding to about 150
 electrons per crystal.
 The mean $\langle C^{\mathrm{mc}} \rangle$ and their r.m.s. values
 $\sigma^{\mathrm{mc}}$ are listed in the Table~\ref{tab4}. In principle,
 the coefficients may depend on the electron energy, layer
 number, and crystal size. But at this
 level of statistics no significant difference in
 $C_{i}^{\mathrm{mc}}$ values for different crystal sizes in the ``large''
 angle part is seen. The statistical accuracy $\sigma_C^{\mathrm{mc}}$ of 
 $C_{i}^{\mathrm{mc}}$ can be estimated as 
 $\sigma_{\mathrm{C}}^{\mathrm{mc}} = \sigma^{\mathrm{mc}} /
 \langle C^{\mathrm{mc}} \rangle$ (~Table~\ref{tab4}~).

 To obtain $C_{i}^{\mathrm{exp}}$ constants,
 $e^+e^- \rightarrow e^+e^-$ events corresponding to integrated luminosity
 about $130~\mathrm{nb}^{-1}$ are needed. Such sample contains about 240000
 electrons in the ``large'' angles part, corresponding to at
 least 150 electrons
 per crystal. On average, the SND acquires such an integrated  luminosity in
 three days of VEPP-2M operation. The mean
 $\langle C^{\mathrm{exp}} \rangle$, their r.m.s. values
 $\sigma^{\mathrm{exp}}$ and  statistical accuracy of
 $C_{i}^{\mathrm{exp}}$ ($\sigma_{C}^{\mathrm{exp}}$) are listed in the
 Table~\ref{tab4} together with the mean $\langle C^{\mathrm{cal}} \rangle$,
 their r.m.s. values $\sigma^{\mathrm{cal}}$ and statistical errors
 $\sigma_{C}^{\mathrm{cal}} =
 \sqrt[]{(\sigma_{C}^{\mathrm{exp}})^2 +
 (\sigma_{C}^{\mathrm{mc}})^2}$. 

 The statistical accuracy of calibration constants for the first two
 layers is satisfactory, but for the third layer it is larger than that 
 for cosmic calibration  \cite{CCC}. This is due to relatively small
 energy deposition of electromagnetic showers in the third layer, combined 
 with large number of hit crystals. Thus the influence of the accuracy of
 calibration constants in the third layer on an overall calorimeter
 resolution for electrons and photons is small.
 The situation is different for muons and charged pions, where relative
 energy deposition in crystals of the third layer is large. In this case
 high statistical error in calibration coefficients significantly increases
 the widths of energy deposition 
 spectra. So, instead  of $C_{i}^{\mathrm{cal}}$, 
 the coefficients obtained during cosmic calibration were
 used  for the third layer.

 The r.m.s. difference $\sigma^{\mathrm{sys}}$ in calibration coefficients
 obtained using cosmic and $e^+e^-$ calibration procedures can be estimated
 as $\sigma^{\mathrm{sys}} =
 \sqrt[]{(\sigma^{\mathrm{exp}})^2 - (\sigma_{C}^{\mathrm{exp}})^2}$.
 This value is about $4\%$ for
 the first two layers and is less than $5\%$ for the third layer.
  
\section{Energy resolution of the calorimeter.
         Implementation of the calibration procedure}

 As a result of $e^+e^-$ calibration the calorimeter energy resolutions
 for 500 MeV electrons and photons (~Table~\ref{tab1},~Fig.3~) were improved
 by $10 \%$, but still remain worse than those expected from Monte Carlo
 simulation.

 A Monte Carlo simulation of energy deposition was first carried out with
 a uniform description of the nonuniformity of the light collection
 efficiency over the crystal volume. Then, calculations taking into
 account nonuniform ``direct'' light from a scintillation and diffuse
 reflection from the crystal boundaries and wrapping were performed.
 The energy deposition $U$ in crystal measured in experiment is

\begin{equation}
 U(\mathrm{MeV}) = C(\mathrm{MeV/pC}) \cdot e \cdot \nu \cdot
 \zeta(\mathrm{1/MeV}) \cdot \psi \cdot E(\mathrm{MeV}),
\end{equation}

 where $E$ is an energy deposition in the scintillation counter, $C$ --
 ratio between the collected electric charge from photodetector, measured
 in pC, and energy deposition in the units of MeV, $\nu$ - quantum efficiency
 of the photodetector multiplied by its gain, $\zeta$ - 
 light yield of the scintillator, $\psi$ - light collection 
 efficiency, e - electron charge. In general, $\psi$ is a function of
 coordinates within the crystal and depends on reflection coefficient on the
 crystal boundaries and photocathode diameter.

 To take into account the nonuniformity of light collection efficiency,
 the crystal response was simulated taking into account a uniform light 
 collection due to diffuse reflection and nonuniform ``direct'' light
 collection, depending on a solid angle of a photocathode, visible from
 a scintillation point. The results of such simulation
 the energy distribution width for 500 MeV photons is
 $4.2 \% (\mathrm{FWHM/2.36})$ (~Fig.~\ref{f4}~). The experimental spectrum
 is shown in the same figure. Calorimeter resolutions for other types of
 particles are also shown in Table~\ref{tab1}. The agreement between
 experiment and  simulation became much better and the residual
 disagreement could be attributed to difference in
 diffuse reflection coefficients values in different crystals.

 To study the calorimeter response for photons as a function of photon
 energy, the events $e^+e^- \rightarrow \gamma\gamma$ and
 $e^+e^- \rightarrow e^+e^-\gamma$ were used. The kinematic fit
 of the $e^+e^- \rightarrow e^+e^-\gamma$ events was
 performed taking into account energy--momentum conservation
 and the reconstructed photon energies $E^{\gamma}_{\mathrm{rec}}$ were
 obtained. These values were compared with direct calorimeter measurements.
 One could expect strong correlation between these values, but study of
 the simulated $e^+e^- \rightarrow e^+e^-\gamma$
 events, where the energy of the photon is precisely
 known, showed, that even for photons in the energy range from 30 up to
 150 MeV these correlation do not change significantly the calorimeter
 response. The energy range above 170 MeV was also studied using $2\gamma$
 annihilation events and results agree well with those for
 $e^+e^- \rightarrow e^+e^-\gamma$ reaction. The dependence
 of calorimeter energy resolution on photon energies (~Fig.~\ref{f9}~)
 was fitted as

\begin{equation}
 \sigma_E/E(\%) = {4.2\% \over \sqrt[4]{E(\mathrm{GeV})}}
\end{equation}

 After calorimeter calibration with $e^+e^- \rightarrow e^+e^-$ events,
 the photon energies turned out to be biased by about $1\%$
 (~Table.~\ref{tab1}, Fig.~\ref{f4}~). In oder to compensate this bias,
 the calibration coefficients for photons were corrected accordingly.
 The distributions over two-photon invariant masses 
 $m_{\gamma\gamma}$ in $\phi \rightarrow \eta \gamma$ and
 $K_s \rightarrow \pi^0 \pi^0$ decays after such correction are shown
 at Fig.~\ref{f5}~and~\ref{f6}.  Peaks
 at $\pi^0$ and $\eta$ mesons masses are clearly seen.

 Relative drift of calibration coefficients is shown in Fig.\ref{f10}.
 It can be seen, that for a time period between consecutive calibrations,
 the mean shift of the coefficients is about $1\%$ and r.m.s. of their
 random spread is about $2.5\%$ for the first two layers and about $5\%$
 for the third layer.

\section{Conclusion}  

 Using the described procedure for the SND calorimeter calibration
 the statistical accuracy of $2\%$ in calibration constants for the
 first two layers was achieved. The final resolution for photons
 varies from $10\%$ at 50 MeV to $5\%$ at 500 MeV.

\newpage

\begin{table}[h]
\caption{Calorimeter response for 500 MeV electrons, photons and muons.
         $E_0 = 500~\mathrm{MeV}$~---~energy of the particles,
         $E^{e,\gamma,\mu}$ --- measured energies in the calorimeter
	 for electrons, photons and energy deposition for muons respectively,
         $\sigma$~---~FWHM/2.36 of distribution over $E/E_0$,
         Peak --- the peak position in the distribution over $E/E_0$.
         EXP1 and EXP2 --- experimental distributions after cosmic and
         $e^+e^-$ calibrations respectively.
         MC1 and MC2 --- distributions in Monte Carlo simulation without and
	 with nonuniformity of light collection over crystal volume was taken
	 into account respectively.}
\label{tab1}
\begin{tabular}[t]{|c|cc|cc|cc|cc|}
\hline
&\multicolumn{2}{c|}{EXP1}&
\multicolumn{2}{c|}{EXP2}&
\multicolumn{2}{c|}{MC1}&
\multicolumn{2}{c|}{MC2} \\ \cline{2-9}
 ~~         & Peak & $\sigma ( \% )$
            & Peak & $\sigma ( \% )$
            & Peak & $\sigma ( \% )$
            & Peak & $\sigma ( \% )$ \\ \hline
 $E^e/E_0$        & 0.99 & 5.4 & 1    & 4.7 & 0.99 & 3.5 & 1    & 4.2 \\ \hline
 $E^{\gamma}/E_0$ & 1    & 5.4 & 1.01 & 5.0 & 1    & 3.7 & 0.99 & 4.2 \\ \hline
 $E^{\mu}/E_0$    & 0.33 & 8   & 0.33 & 8   & 0.33 & 5   & 0.34 & 7   \\ \hline
\end{tabular}
\end{table}

\begin{table}
\caption{$C^{\mathrm{mc}},~C^{\mathrm{exp}},~C^{\mathrm{cal}}$ coefficients
         and their statistical accuracy in calorimeter layers for crystals
         at the "large" angle zone}
\label{tab4}
\begin{tabular}[t]{|c|cc|c|cc|c|cc|c|}
\hline
 layer number
 & $\langle C^{\mathrm{mc}} \rangle$ & $\sigma^{\mathrm{mc}}$ &
 $\sigma_{\mathrm{C}}^{\mathrm{mc}}(\%)$ 
 & $\langle C^{\mathrm{exp}} \rangle$ & $\sigma^{\mathrm{exp}}$ &
 $\sigma_{\mathrm{C}}^{\mathrm{exp}}(\%)$ 
 & $\langle C^{\mathrm{cal}} \rangle$ & $\sigma^{\mathrm{cal}}$ &
 $\sigma_{\mathrm{C}}^{\mathrm{cal}}(\%)$ \\ \hline
 I   & 1.04 & 0.02 & 1.9 & 1.07 & 0.04 & 1.5 & 1.03 & 0.04 & 2.3 \\ \hline
 II  & 1.05 & 0.02 & 1.8 & 1.05 & 0.04 & 1.5 & 1    & 0.04 & 2.3 \\ \hline
 III & 1.29 & 0.06 & 5   & 1.27 & 0.07 & 4   & 0.98 & 0.08 & 6.4 \\ \hline
\end{tabular}
\end{table}

\clearpage

\begin{figure}
\epsfig{figure=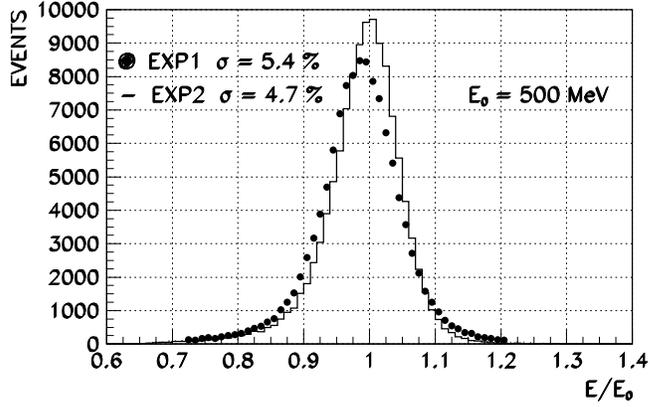,height=7cm}
\caption{Energy spectra for 500 MeV electrons;
         $E_0 = 500$ MeV - beam energy,
         $E$ - measured energy.
         EXP2 - distribution after $e^+e^-$ calibration,
         EXP1 - distribution after cosmic calibration.}
\label{f3}
\end{figure}

\begin{figure}
\epsfig{figure=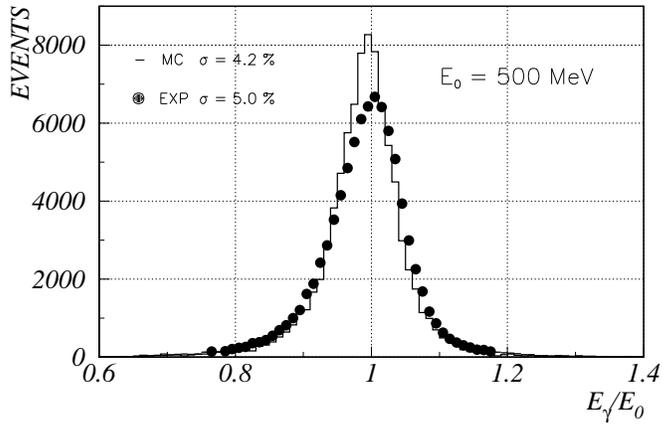,height=7cm}
\caption{Energy spectra for 500 MeV photons. $E_0 = 500$ MeV - beam energy,
         $E_{\gamma}$ - measured energy.}
\label{f4}
\end{figure}

\begin{figure}
\epsfig{figure=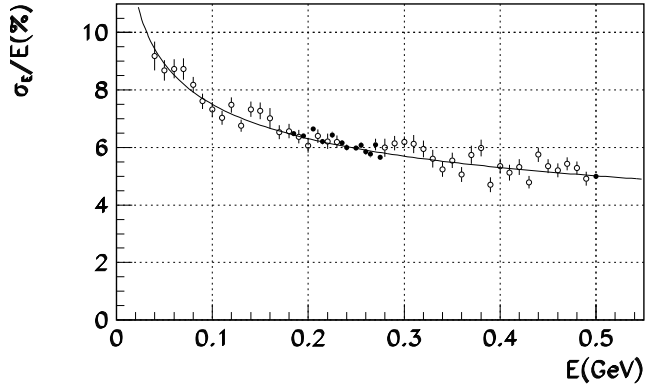,height=7cm}
\caption{Dependence of the calorimeter energy resolution on
         the photon energy. $E$ - photon energy,  $\sigma_E/E$ - energy
         resolution of the calorimeter obtained using 
         $e^+e^- \rightarrow \gamma\gamma$ ( dots ) and
         $e^+e^- \rightarrow e^+e^-\gamma$ ( circles )reactions. The error
	 bars show only statistical errors.}
\label{f9}
\end{figure}

\begin{figure}
\epsfig{figure=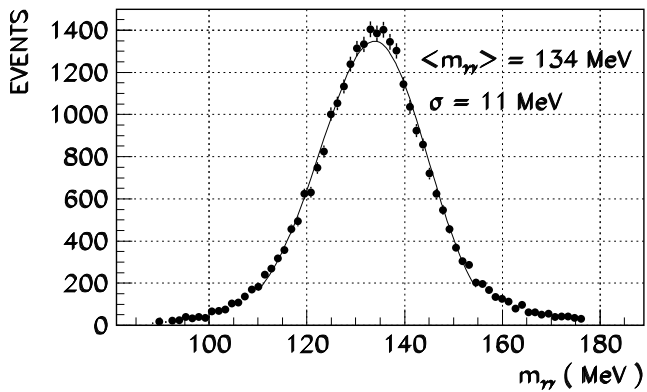,height=7cm}
\caption{Two photon invariant mass distribution in the experimental
         $\phi \rightarrow K_S K_L$, $K_S \rightarrow \pi^0 \pi^0$ events.
         Line - asymmetric Gaussian fit.}
\label{f5}
\end{figure}

\begin{figure}
\epsfig{figure=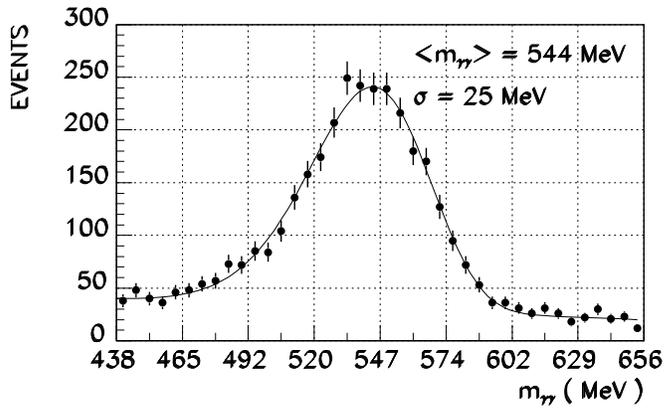,height=7cm}
\caption{Two photon invariant mass distribution in experimental
         $\phi \rightarrow \eta \gamma$ events.
	 Line - asymmetric Gaussian fit.}
\label{f6}
\end{figure}

\begin{figure}
\epsfig{figure=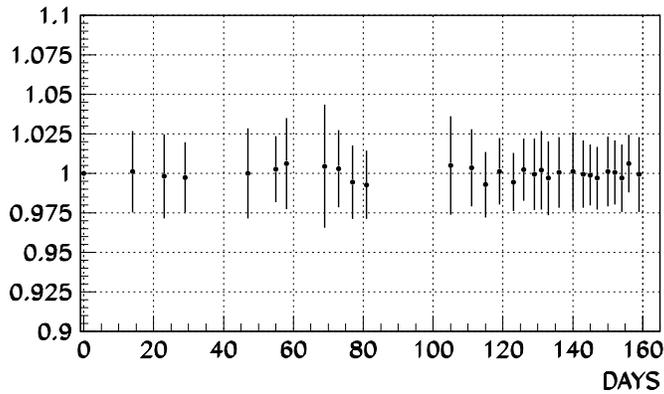,height=7cm}
\caption{The calibration coefficients spread. Points --- average ratio of
         a current calibration result to the preceding one, error
	 bars --- FWHM/2.36 of the distributions of these ratios over
         the whole calorimeter layer. Horizontal axis shows the time elapsed
	 from the first calibration. Shown are the results for the second
	 calorimeter layer. Other layers behave similarly.}
\label{f10}
\end{figure}

\end{document}